\label{sec_conc_4}%
\documentclass[%
 reprint,
 amsmath,amssymb,
 aps,
]{revtex4-1}

\usepackage{graphicx}% Include figure files
\usepackage{dcolumn}% Align table columns on decimal point
\usepackage{bm}% bold math
\usepackage{xcolor, soul}
\usepackage{textgreek}
\usepackage{fancyhdr}
\begin{document}

\title{Experimental study of second sound quench detection for superconducting cavities}

\author{J. Plouin}
 \email{juliette.plouin@cea.fr}
 
\author{B. Baudouy}
\author{A. Four}
\author{J.P. Charrier}
\author{L. Maurice}
\author{J. Novo}
\affiliation{%
 {CEA Paris-Saclay, Universit\'e Paris-Saclay, Gif-sur-Yvette, France}
}%

\author{B. J. Peters}
\affiliation{
Karlsruher Institut fur Technologie Postfach 3640 76021 Karlsruhe, Germany
}%

\author{K. Liao}
\affiliation{
 Imperial College London and Ideabatic LTD, SW7 2AZ, 32 Lingrey Court, Cambridge, 10 CB2 9JA, United Kingdom
}%
\date{Received 28 February 2019; published 26 August 2019}

\begin{abstract}
Superconducting RF cavities are used in particle accelerators to provide energy to the particle beam. Such cavities are mostly fabricated in niobium and often operated in superfluid helium. One of their limits of operation is the appearance of a local quench, initiated by a local field enhancement due to a defect, which leads to a normal conducting transition of the cavity. Localizing the quench area can be achieved with temperature mapping systems. Another method is the use of second sound wave propagation in superfluid helium. Measuring the time of propagation of these waves from quench location to special sensors, called Oscillating Superleak Transducers (OSTs), and using their well-known velocity should allow trilateration. However, most of experimental measurements on cavities show “premature signals”, i.e. the second sound signals arrive earlier on the OSTs than expected. 
This paper presents several quench experiments on cavities equipped with OSTs and temperature mapping quench detection systems. Two hypotheses can explain the observed premature signals. The first one assesses faster propagation in helium. An experimental setup has been developed for testing this hypothesis, where second sound is created by a localized heater in a controlled environment up to 4.3~kW/cm$^2$ and 2.8~J. Premature signals could not be verified in this setup. A second hypothesis based on a simple model including several processes in niobium and second sound propagation in helium is discussed. The model improves significantly the prediction of the times of arrival of the second sound waves. The overall study shows that the processes in niobium play a prominent role in the second sound detection for superconducting cavities.

\begin{description}
\item[Keywords]
Superconducting cavity, Quench, Superfluid helium, Second sound
\end{description}
\bigskip
\large{Published in Physical Review Accelerators and Beams 22, 083202 (2019)\\ doi: 10.1103/PhysRevAccelBeams.22.083202}
\end{abstract}

\maketitle
\thispagestyle{fancy}

%\tableofcontents

\section{Introduction}

Radio-frequency cavities are state of the art technology in high energy particle accelerators to increase the momentum of charged particles. The resonating electromagnetic field causes coulomb heating on the inner surfaces of these cavities. Normal-conducting cavities are usually made of copper. The generated heat can be greatly deceased by using a superconducting cavity in the Meissner state. The element with the highest lower critical field is niobium, which is thus mostly used for this purpose \cite{Padam2008}. To generate high accelerating gradients, the cavities are operated close to their operational limit. A number of different processes are known to limit the operation of superconducting cavities \cite{Padam2008}. The limiting process that normally occurs for very well performing cavities is the localized quench where a local increase of the electromagnetic field due to a defect drives a local area to normal-conducting transition, and due to the increased heating to the global normal-conducting state \cite{Geng2009}. Even small imperfections on the inner cavity surface can cause this quench. These spots can be treated once their locations are detected. A number of methods has been applied for this \cite{Iwashita2014, Navitski2014}.

Since the 1980’s temperature sensors are used to detect the temperature increase on the outer cavity surface \cite{Conway2017}. These temperature mapping systems are based on temperature-dependent carbon sensors, mostly Allen-Bradley resistors \cite{Padam1987}. In addition to being sensitive for temperature measurements in both normal (He~I) and superfluid (He~II) helium, their geometry can be customized to match the topology of the superconducting cavity. Some tests were made with temperature sensors that were glued to the cavities surface, but mechanical systems that press the sensor to the cavity surface are more common. The glued sensors are normally not reusable and a high number of sensors are presented and thus read-out systems are needed. The mechanical systems require custom-fit frame for each cavity geometry, but for elliptical cavities they can be mounted on a rotating frame, that allows moving them around the cavity in order to locate the quench spot. Another method for the detection is the utilization of the special heat transport properties in He~II. In superfluid helium, heat is transported as a wave, called second sound. This wave travels with a known, temperature-dependent velocity. If at least three sensors, capable of detecting second sound with precise time information and known positions relative to the cavity were mounted, trilateration of the quench spot is possible. Conway et al. introduced this method and utilized oscillating superleak transducers (OSTs) for the detection of second sound waves \cite{Conway2008}.

For the last ten years, quench localization with 2\textsuperscript{nd} sound detection has been developed in several labs. It was found that this method is cheaper and easier to implement than temperature mapping systems; however, nearly all these laboratories have observed the same kind of premature signals: the second sound signal arrives on the OSTs earlier than expected \cite{Maximenko2011, Liao2012, Plouin2013, Tama2016, Junginger2015, Koettig2015}. Efforts have been pursued to explain these premature signals and hypothesis have been envisaged. Eichhorn et al. \cite{Eichhorn2015} have supposed that the second sound wave could be generated before the RF system can detect the quench; they have developed a corresponding model, which could explain some tens of \textmu s error, much lower than experimentally observed. Liu \cite{Liu2012} and Eichhorn \cite{Eichhorn2015b} have considered the heat diffusion in niobium, in addition to the second sound propagation, which could cause premature signals, but their models have not been systematically compared to experimental results. The possibility of other effects beside second sound propagation in He~II has also been considered, and experiments have been carried out with local heaters as a source for the heat propagation. Liao et al. \cite{Liao2011} have generated heat flux up to 200 W/cm$^2$, and did not observe premature signal, while Junginger et al. \cite{Junginger2015}, with heat flux up to 350 W/cm$^2$ observed 10\% faster signals in particular conditions. Very recently, a new technique using flow visualization in superfluid helium has been proposed by Bao et al. to detect quench-spot in RF cavities  \cite{Bao2019}. A prof-of-concept experiment with a miniature heater is presented for heat flows below 300 W/m$^2$. A ideal boiling zone model is proposed to explain the premature signal on the data obtained by \cite{Junginger2015} and \citep{Koettig2015}, but no test with a RF cavity has been performed yet.

Up to now, none of these hypothesis and associated models could properly describe premature signals measured during cavity tests.

The present paper is a contribution to these efforts. We performed five tests on two single-cell superconducting cavities, with variation of the helium bath temperature for one of them, and always observed premature signals. In order to study the pure heat propagation in He~II, we carried out experiments with a localized heater as a source for the second sound wave. We reached values of surface power (4.3~kW/cm$^2$) and global energy (2.8~J) that were never achieved in similar experiments, for these experiments no premature signal was observed. 

\begin{figure}[h]
\centering
\includegraphics[width=8.6cm]{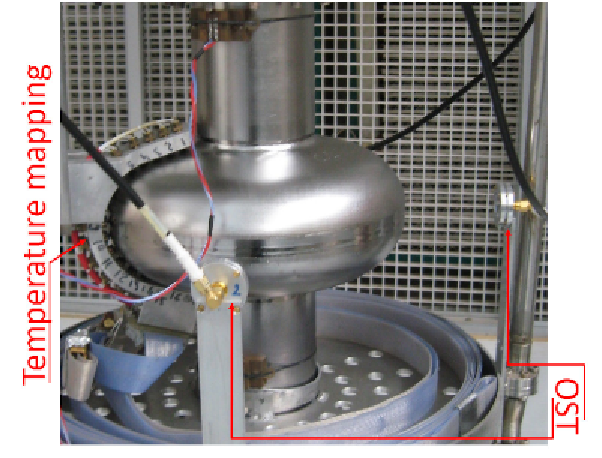}
\caption{1.3~GHz single-cell cavity equipped with temperature mapping system and OSTs.}
\label{cav-setup}
\end{figure}

\section{Tests on Cavities}
\subsection{Setup and cavities test description}

At CEA Paris-Saclay, superconducting cavities are usually tested in superfluid helium in a dedicated vertical cryostat on the Supratech platform \cite{Cenni2017, Baudouy2010}. For several single-cell cavities such tests could be realized with second sound sensors for quench localization. During those tests, each cavity was equipped with a temperature mapping system, and with four OSTs facing the cavity in the equator plane, placed at around 6.5~cm from the cavity. All these elements can be seen in Figure \ref{cav-setup}. Two Tesla shape single-cell cavities, C1-21 \cite{Eozenou2010, Eozenou2012} and 1AC3 \cite{Eozenou2014} have undergone these tests; they will be named respectively A and B. These cavities have the same elliptical geometry, and thus identical RF parameters: frequency f~=~1.3~GHz, shunt impedance r/Q~=~104~\textOmega and geometrical factor G~=~283~\textOmega. They were supplied by two different manufacturers, from niobium plates with RRR values between 250 and 300, and both of them have undergone several chemical and electropolishing treatments.

\begin{figure}[h]
\centering
\includegraphics[width=8.6cm]{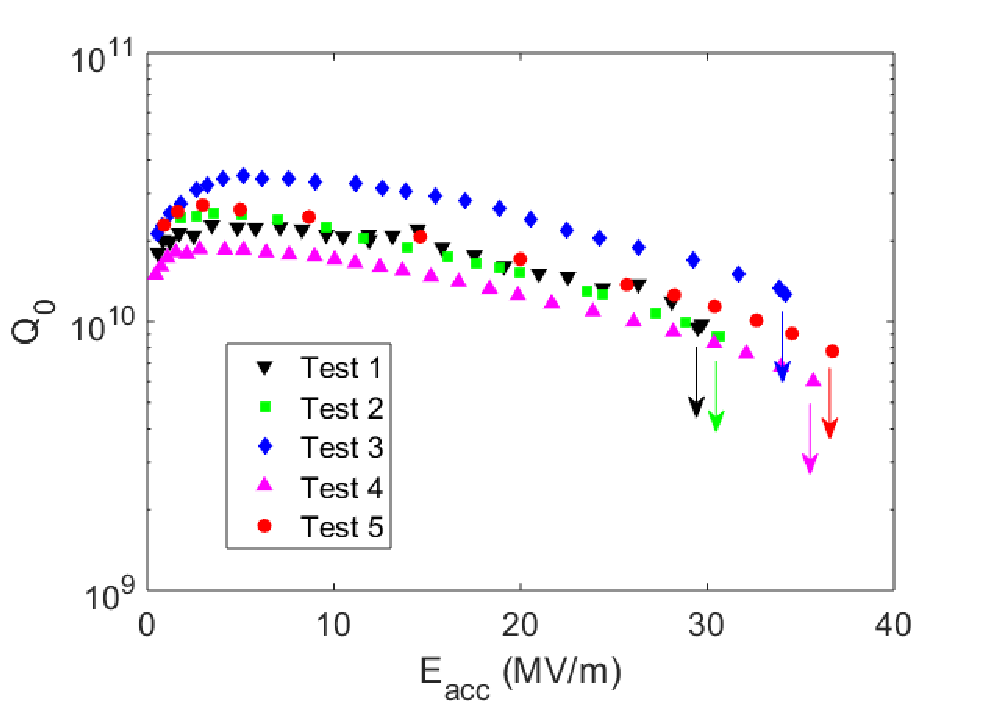}
\caption{$Q_0/E_{acc}$ curves of the tested single-cell cavities. Vertical arrows show quench appearance}
\label{Q-Eacc}
\end{figure}

Five tests where localized quench was reached without electron emission could be realized: two tests with Cavity A and three tests with Cavity B. The $Q_0$ vs. $E_{acc}$ curves are shown in Figure \ref{Q-Eacc}, $Q_0$ being the cavity quality factor and $E_{acc}$ the accelerating field. The accelerating field is increased by injecting RF power in the cavity, until the cavity quenches, which means that the injected power is entirely reflected. The quench appearance is indicated by a vertical arrow for each test, which thus determines the maximum accelerating field allowed in the cavity ($E_{max}$). The test results are summarized in Table \ref{tests}. The differences in the $Q_0$/$E_{acc}$ curve for Cavity B from one test to the another is due to the intermediate chemical/electropolishing treatments. The helium bath temperature was about 1.5~K for Cavity A and about 1.6~K for Cavity B.

\begin{table}[]
    \centering
    \caption{ Cavity tests summary}
    \begin{tabular}{>{}c >{}c >{}c}
            & Cavity name & $E_{max}$\\
            \hline
    test 1    &  A & 30 MV/m  \\
    test 2    &  A & 31 MV/m  \\
    test 3    &  B & 34 MV/m  \\
    test 4    &  B & 35 MV/m  \\
    test 5    &  B & 37 MV/m  \\
    \end{tabular}
 
    \label{tests}
\end{table}

\subsection{Quench localization with temperature mapping system}

\begin{figure}[h]
\centering
\includegraphics[width=5cm]{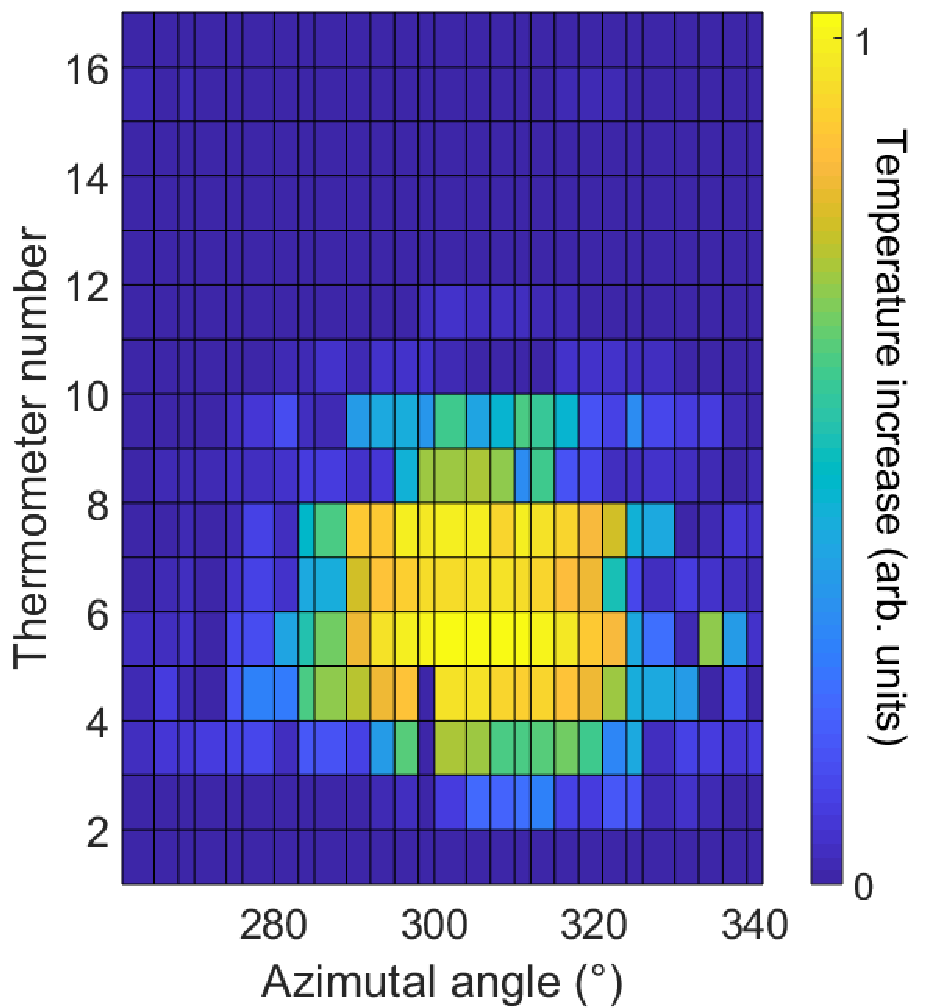}\\
\includegraphics[width=4.3cm]{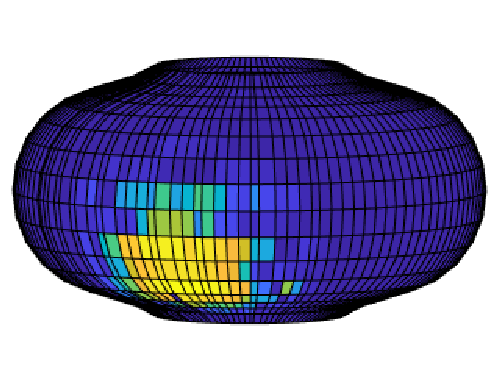}\hfill
\includegraphics[width=4.3cm]{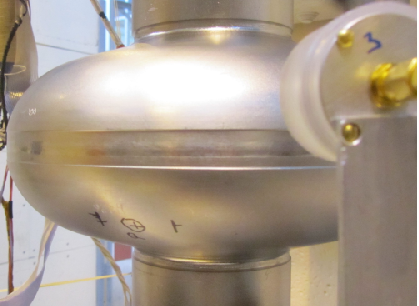}
\caption{Top: Thermal cartography of Cavity B during a test. Azimuthal angle gives the temperature mapping position around the cavity; thermometer number can be read in Figure~\ref{cav-setup}. Bottom: Cartography projection on cavity surface (left) and identification of the quench spot on the cavity marked by $\oplus$ symbol (right).}
\label{carto}
\end{figure}

Initially the quench position was localized by the temperature mapping system, in order to have a reference for comparison with the OSTs measurements. This system consists of 17 thermometers fixed on a structure able to rotate around the cavity axis \cite{Berry2003}. Each thermometer is constituted of a carbon resistor placed inside a copper support which is in contact with the cavity surface. The aim of this measurement is to observe the temperature elevation when the quench occurs. The system works as follows: at each azimuthal position, the 17 thermometers are successively monitored. The duration of a single readout is long compared to the quench dynamics, so the following method is used: the sequence is repeated about twenty times while the cavity is forced to quench. For each thermometer the maximum value of this series of measurements is used, because it corresponds to the closest in time to the quench occurrence. As an example the temperature array is shown in Figure~\ref{carto} around the quench location (top), and is projected on the cavity surface (bottom). The resolution is about~3~degrees in latitude and 1~cm in longitude (distance between two sensors).

This temperature mapping system is used to produce a map showing the measured temperature with respect to the bath. As the mapping system itself is immersed in the bath, the temperature given by the carbon resistors is different from the surface cavity temperature. A temperature difference equal to zero means that locally the cavity surface is at the helium bath temperature. This temperature mapping system is efficient to achieve a localization within a precision of a less than one centimeter in each direction. According to the temperature maps, the quench location of Cavity A always lands at the equator and that of Cavity B at the bottom half of the cavity (see Figure~\ref{carto}).

\subsection{OST measurements at 1.7 K}
An Oscillating Superleak Transducer (OST) is a sensitive sensor to detect second sound. The idea for this device originated in 1968 \cite{Williams1968}. It consists of a capacitor like setup with one flexible membrane and a solid back electrode. The membrane is flexible and has nanometer-sized pores. For the shown measurements a metal-coated membrane with holes of about 100~nm were used. According to the two-fluid theory the superfluid component of the helium can pass through these holes while the normal fluid component cannot \cite{Donnelly2009}. Thus, the arrival of a second sound wave, that has different amounts of the two components in comparison to the helium between the two plates \cite{Donnelly2009}, applies a pressure on the flexible membrane, decreasing the distance between the two membrane plates. The two sides are charged with 120~V via a weakly coupled connection. For fast processes, such as a second sound arrival, the charge on the OST stays nearly constant. A decrease of the distance of the capacitor plates thus causes a drop in voltage of the membrane \cite{Quadt2012}. Each OST has its own electronic circuit which provides the polarization and amplification of the signal. For all the measurements presented in this paper, the same set of OSTs, with diameter 20~mm, was used. This set was provided by Cornell University \cite{Conway2009}.

\begin{figure}[h]
\centering
\includegraphics[width=8.6cm]{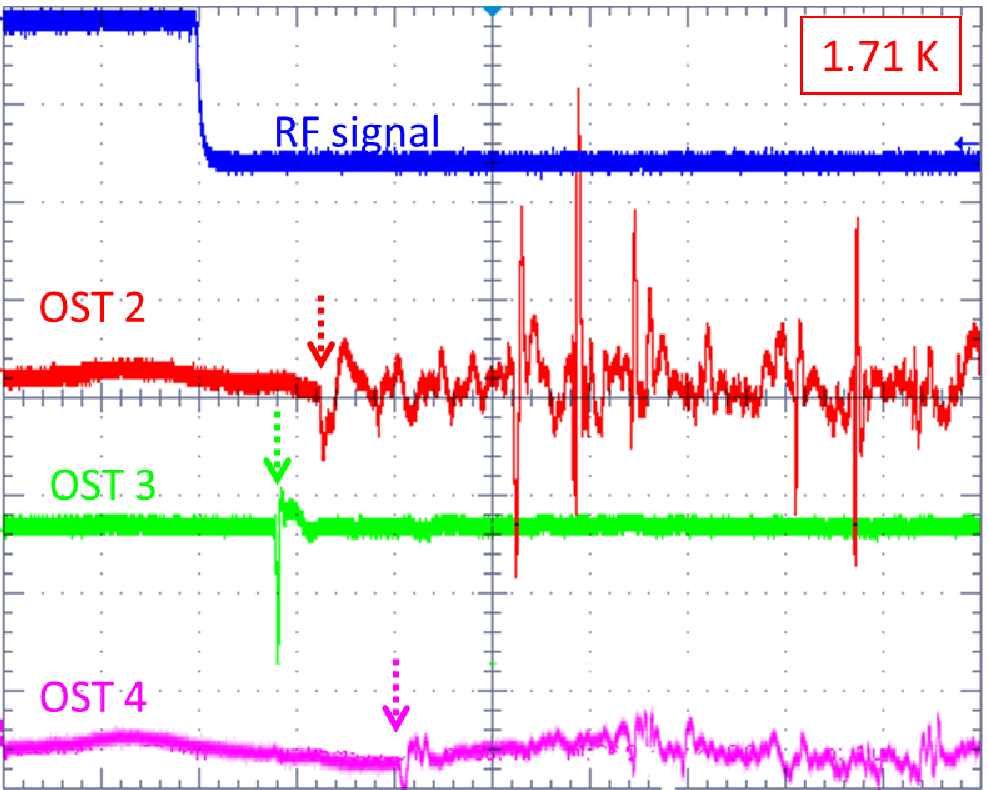}
\caption{Oscilloscope signals from OSTs during cavity quench. Vertical arrows show 2\textsuperscript{nd} sound wave arrival on OST.}
\label{T171}
\end{figure}

For each test, the signals from OSTs during a series of quenches were measured. The results were recorded by an oscilloscope that was triggered by the reflected RF power from the cavity, indicating a quench. A typical oscilloscope signal is shown in Figure \ref{T171}. The blue curve shows the RF signal. As explained previously, the signals from OSTs always start with a sharp negative decrease. The times of arrival of the wave on each OST after quench trigger can be measured with a precision better than 100~\textmu s, and are indicated by vertical arrows on the figure. The OSTs are ordered by their distance from the quench spot, starting with the 1\textsuperscript{st} OST as the closest. During each test, the two nearest OSTs (see Figure \ref{T171}) were nearly facing the quench location, and they presented a clear signal while the two others were hidden by the cavity. The signal of the furthest one was not detectable and is not shown on the figure.

\begin{figure}[h]
\centering
\includegraphics[width=8.6cm]{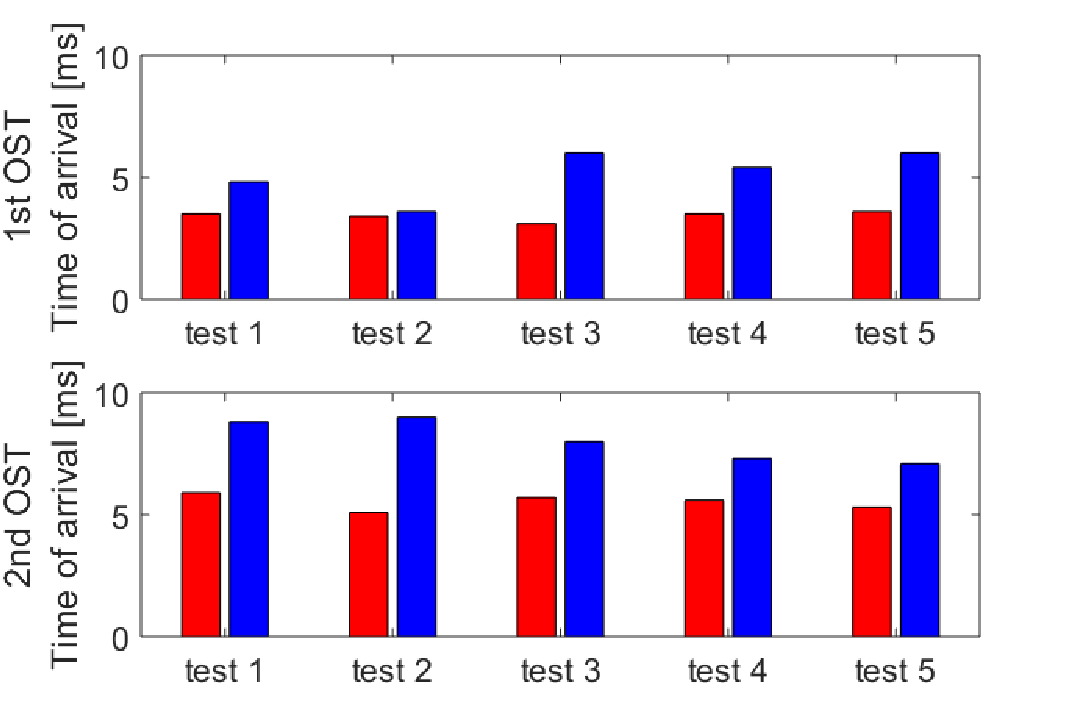}
\caption{Times of arrival of the 2\textsuperscript{nd} sound signal on the two closest OSTs from the quench location at 1.7~K: as measured during experiment (Red) and as calculated with in direct line of sight (Blue).}
\label{bar1}
\end{figure}

The times of arrival of the second sound signal on the two nearest OSTs are represented in Figure \ref{bar1} together with the corresponding times of arrival for the following hypothesis: a pure second sound wave is emitted by a point-like heat source situated at the quench location. These times are thus equal to the distance between OST and quench location (given by temperature mapping system) divided by 20~m/s, which is the second sound velocity at 1.7~K \cite{Donnelly}. In the following, we will refer to this calculation approach as the "direct line of sight".

The difference between measured and calculated times of arrival shows that it is not possible to consider the quench as a punctual emission of a 2\textsuperscript{nd} sound wave: the heat wave reaches the OSTs faster than expected by the direct line of sight propagation. 

Such premature signals have been experimentally observed by nearly all the laboratories which have used second sound sensors to localize a quench in a superconducting cavity \cite{Maximenko2011, Plouin2013, Tama2016, Eichhorn2015, Liu2012}. Different hypothesis have been envisaged to explain this difference, and will be discussed later in this paper.

\subsection{OST Measurements with varying helium temperature}

\begin{figure}[h]
\centering
\includegraphics[width=8.6cm]{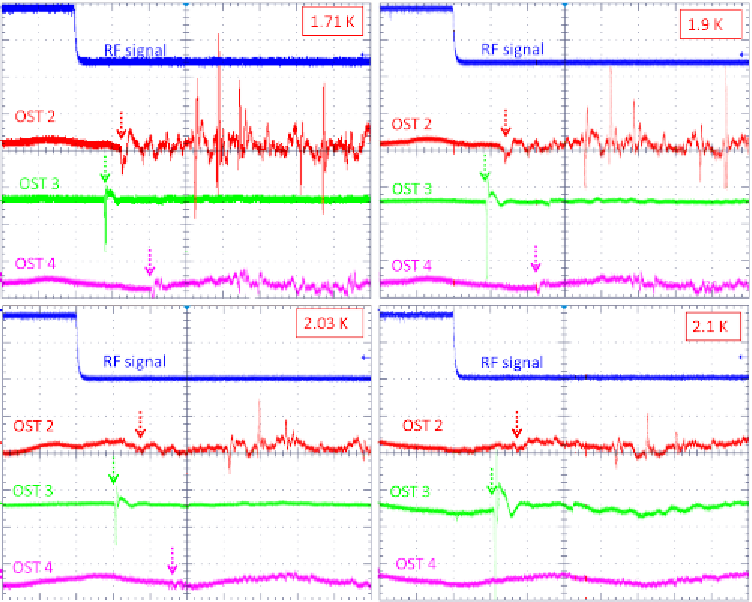}
\caption{Oscilloscope signals from OSTs during cavity quench at different temperatures. Horizontal: 4~ms/division. Vertical. RF Signal: 10m~V/div; 2nd~and~3rd~OST : 100~mV/div; 1st~OST: 1~V/div. (1.71~K), 500mV/div. (1.9, 2.03~K), 100~mV/div (2.1~K).  }
\label{scopevarT}
\end{figure}

During test~4 on cavity B, the helium bath temperature was varied while measuring the evolution of the OST signals during cavity quenches. The scope signals are shown on the Figure \ref{scopevarT} for four temperature values (1.71~K, 1.9~K, 2.03~K and 2.1~K), and for the three closest OSTs.

The dotted vertical arrows show for each OST the time position taken into account for the arrival of the 2\textsuperscript{nd} sound wave. While this position always stays easily detectable for the 1\textsuperscript{st}~OST (the closest to the quench location), one can see that for the other OSTs, the signal becomes less evident to detect with increasing temperature, and is even undetectable at 2.1~K for the 3\textsuperscript{rd}~OST.

\begin{figure}[h]
\centering
\includegraphics[width=8.6cm]{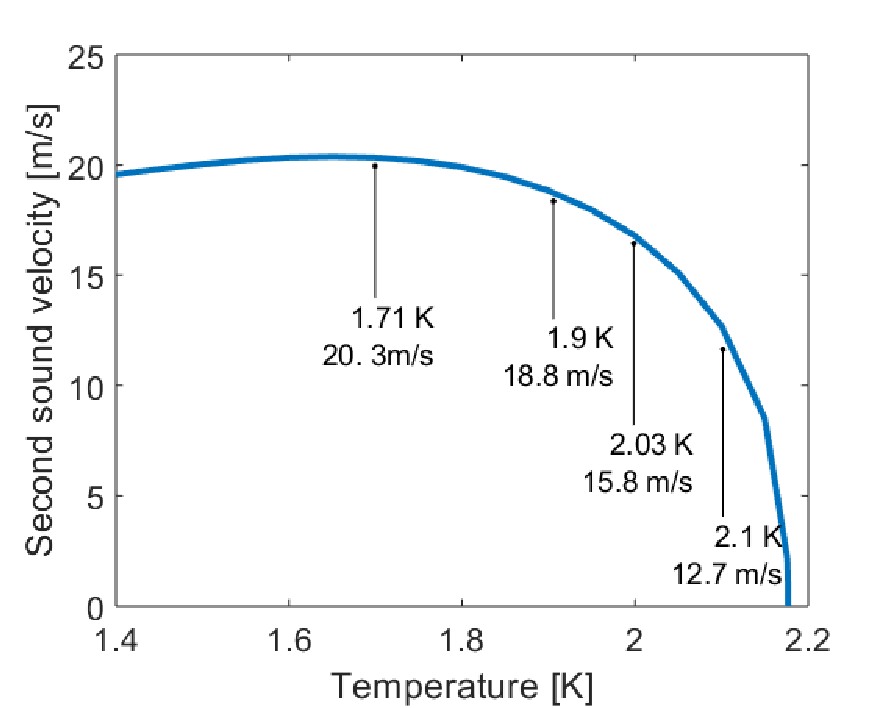}
\caption{Recommended values of the second sound velocity at different temperatures and values relevant for presented experiments, according to \cite{Donnelly}.} 
\label{v_2vsT}
\end{figure}

\begin{figure}[h]
\centering
\includegraphics[width=8.6cm]{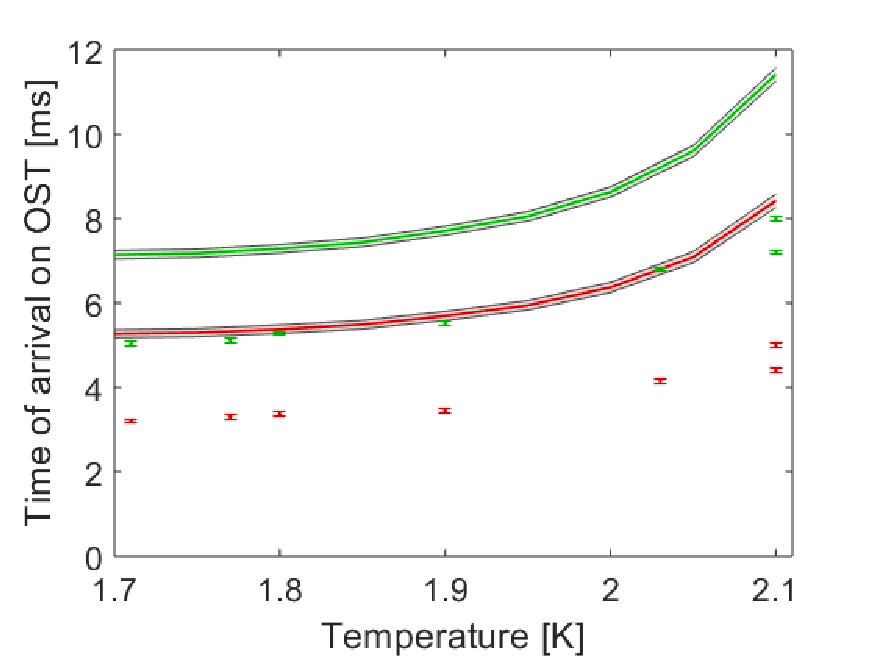}
\caption{Time of arrival on first (red) and second (green) OST with varying temperature: experimental data (points with error bars) and calculated data from the direct line of sight propagation (solid line). The shaded area represents the error on calculation due to geometrical measurements uncertainties.}
\label{TvarOST}
\end{figure}

The experimental results (see Figure \ref{TvarOST}) show how the time of arrival increases with temperature, which is consistent with the evolution of second sound velocity, decreasing with temperature in the [1.7-2.1~K] range, see Fig.\ref{v_2vsT}. For the two closest OSTs, the experimental data are compared to the corresponding times of arrival from the direct line of sight propagation, using the second sound velocity values between 1.7 and 2.1~K \cite{Donnelly}. These results show that the experimental data has the same behavior as the model data when temperature is growing. However, in the whole temperature range, the heat wave reaches the OSTs faster than expected by the direct line of sight propagation. The error bars indicated in Figure~\ref{TvarOST} take into account the uncertainty on the time of arrival measurement ($\pm$ 0.05~ms, error bars on points) and the uncertainty on the distance measurement between the quench spot on the OSTs ($\pm$ 2~mm, shaded area on lines).

\subsection{Conclusions on OSTs measurement during cavity tests}

The first approach considered for the quench localization with OSTs was based on the hypothesis of an instantaneously emitted heat wave by a point like source (the quench location) which travels from this source to the OSTs with the second sound velocity in superfluid helium. This hypothesis is not compatible with the experimental results during cavity quenches in the temperature range [1.7~-2.1~K].

\begin{figure}[h]
\centering
\includegraphics[width=8.6cm]{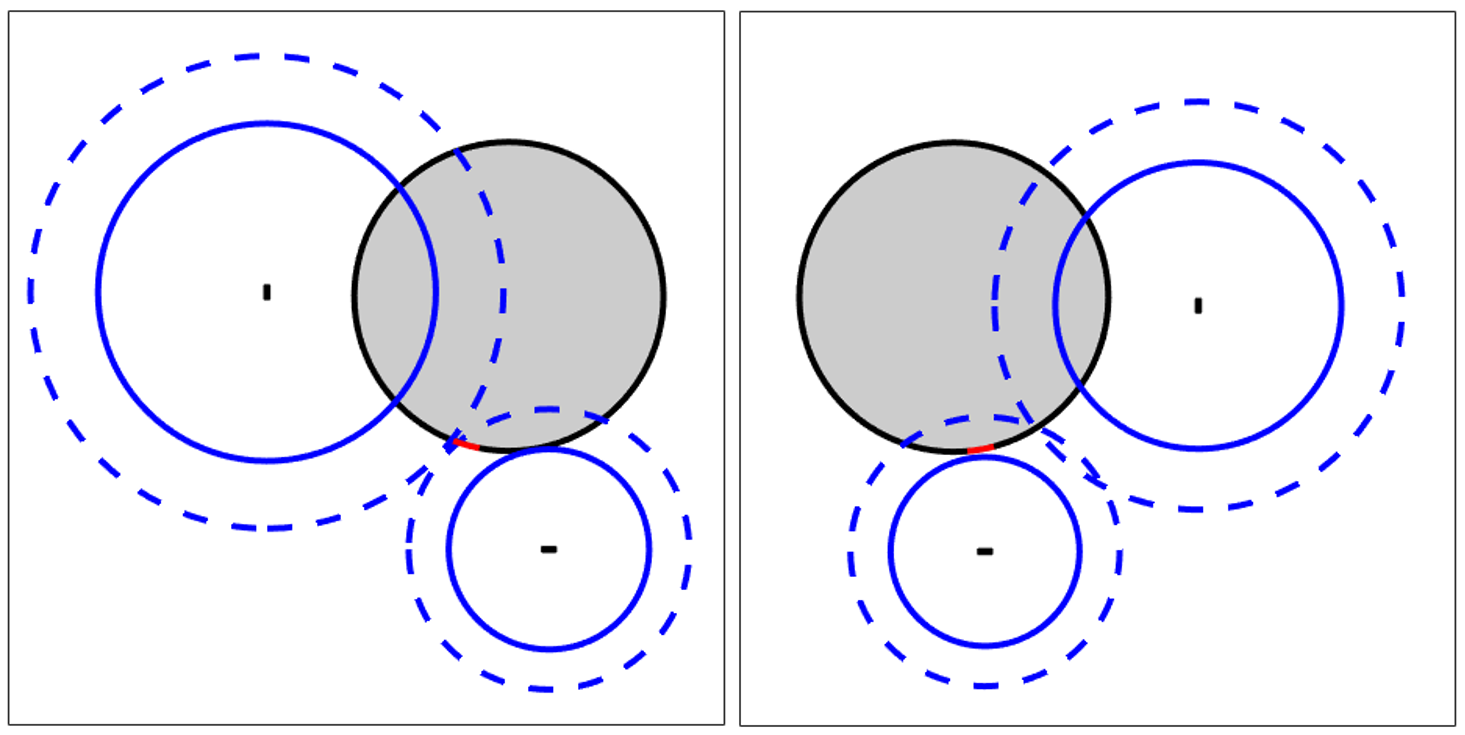}
\caption{2D example of trilateration, for direct line of sight propagation for test~1 (left) and test~2 (right). Cavity is represented seen from above by a grey disk bordered by dark circle corresponding to its equator. Blue plain circles have radii equal to the time propagation for corresponding OST multiplied by 20~m/s. Dashed circles have radii equal to the time propagation for corresponding OST multiplied by 28~m/s. OSTs are represented by small dark rectangles, and quench position detected by temperature mapping system is in red.}
\label{triang}
\end{figure}

Thus, a quench location was unable to be determined with the standard trilateration method. This is illustrated in Figure \ref{triang} for two cases (test~1 and~2) where the problem can be represented in 2D because quench occurs on the equator (and  OSTs are in the equator plane). The blue plain circles have radii equal to the time propagation to corresponding OST multiplied by 20~m/s, the second sound velocity at test temperature. The circles do not intersect, showing that the localization is not possible. When the propagation velocity is arbitrarily increased, up to 28~m/s in those cases, it becomes possible to have circles intersection located on cavity equator (dashed circles). However it does not fit with the quench position measured by temperature mapping system, identified by a red region on equator as shown on the Figure \ref{triang}. This situation becomes even worse if more than two OSTs are used. 

To be able to allow more information sources and account for measurement errors, DESY developed an elegant method \cite{Tama2016}. They include the information, that the heat spot will lie on the surface of the cavity. From each surface point they calculate the length of the shortest path in helium to each OST. Together with the second sound velocity and the OST signal they obtain an expected distance the second sound has travelled in helium. The distance between this previous calculated propagation length $s_{z,\phi}$ and expected distance $d$ can be used for every spot on the cavity surface and $n$-OSTs to determine the root mean square deviation ${RSME}_{z,\phi}$. The area where this deviation is smallest, can be seen as the area where the second sound was triggered.

\begin{equation}
{RMSE}_{z,\phi}=\sqrt{\frac{\sum_{i=1}^{n}(s_{z,\phi}-d_i)^2}{n}}
\end{equation}

This method is used today by DESY to localize the quench position, and comparison with temperature method shows that it works within a few millimeters, if used in the right conditions. In practice, this method helps to localize a small area which can then be examined with some imaging system to find and cure the defect. However, the calculation results given by this method still lead to premature signals.

Discarding the first hypothesis of pure second sound direct line of sight propagation, two new hypothesis can then been envisaged:
\begin{itemize}
    \item Considering another phenomena in helium making the overall thermal signal traveling faster than a pure second sound wave
    \item Considering a part of thermal signal diffusing in niobium making the overall thermal signal traveling faster than a pure second sound wave
\end{itemize}

\section{Propagation in Helium}

In this section the validity of the first hypothesis is explored. 

\subsection{Tests with a localized heater in liquid helium}

A number of ideas where brought forward to explain that the information of the heat input into helium II could be transported faster than the second sound velocity. Temperley theoretically predicted a faster propagation for high heat inputs \cite{Temperley1951}. The theory was refined by Khalatnikov \cite{Khalatnikov1952} and later experimentally proven by Dessler and Fairbank \cite{Dessler1956}. Additionally the appearance of boiling effects could also lead to a faster thermal transport. Either by the emission of a first sound wave, which is approximately ten times faster than second sound at the temperatures in question \cite{Donnelly}, or by effects concerning the gas bubble itself. Although the signals in the cavity tests arrive too late to be seen as pure first sound waves, an unproven theory by Dessler predicted an interaction between first and second sound, which could explain a faster arrival of a second sound wave \cite{Dessler1959}. The expansion of a gas bubble at the heating spot could also be faster than the second sound.

Since the premature signals were observed experimentally, an attempt was made to replicate them in a controlled environment. In order to exclude effects by the cavity itself, the heating was carried out by a small heater, whose output power and energy are regulated.

\subsection{Heat pulse generation \label{Hpg}}
\begin{figure}[h]
\centering
\includegraphics[width=8.6cm]{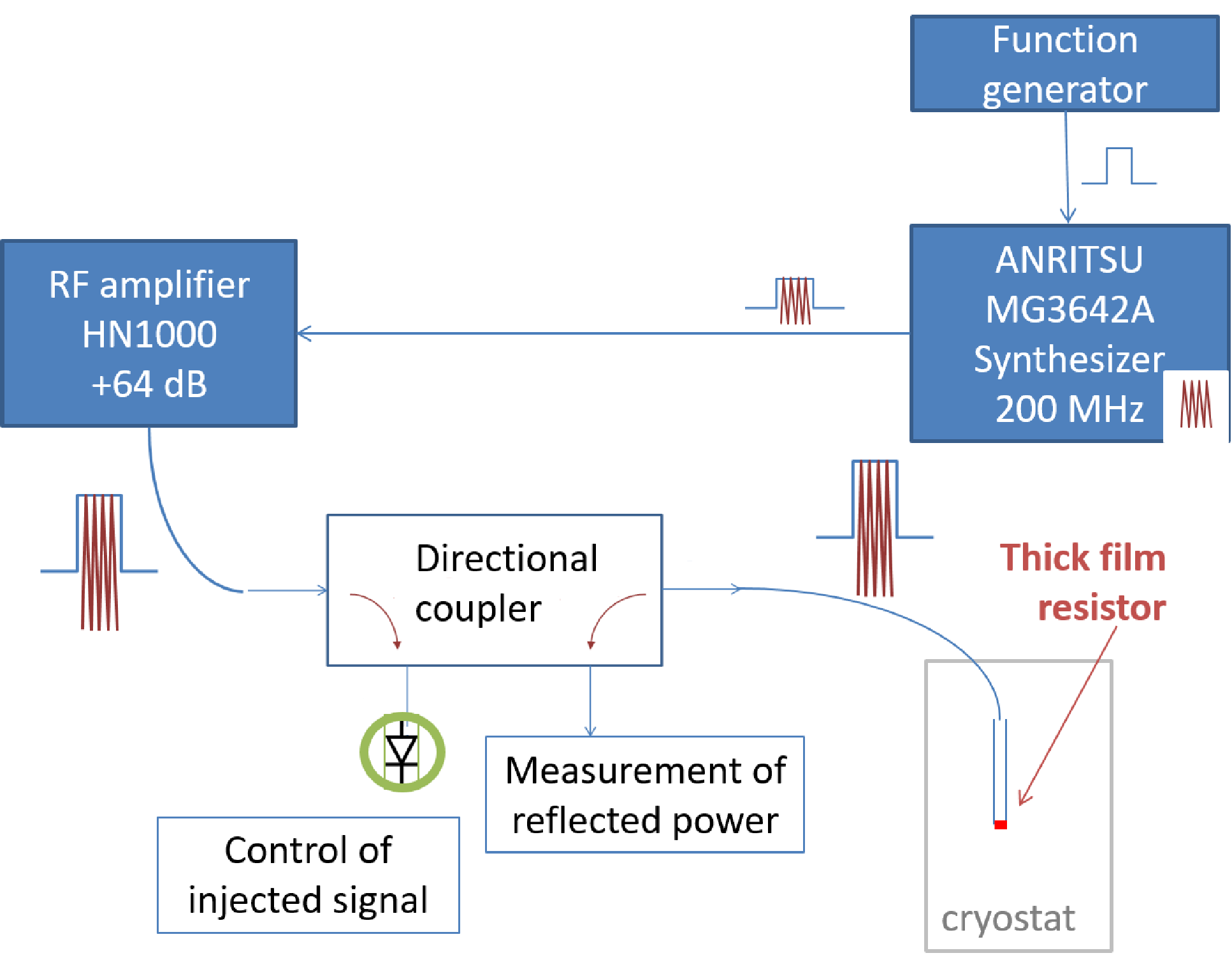}
\caption{Experimental setup for the generation and control of the heat pulse.}
\label{gene}
\end{figure}

\begin{figure}[h]
\centering
\includegraphics[width=8.6cm]{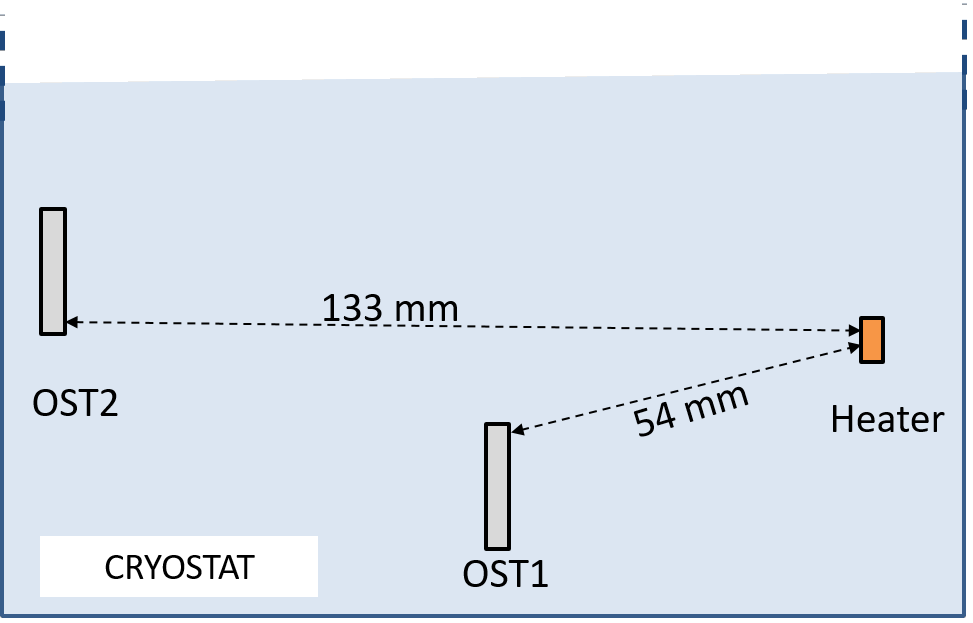}
\caption{Picture of the experimental setup.}
\label{setup}
\end{figure}

 The heat pulses are generated with a 50~\textOmega RuO$_2$-ZnO thick film resistor by the company Bourns, which is specified to accept up to 40~W in DC at room temperature. In the following it was used at cryogenic temperatures for pulsed heat loads of more than 1~kW. This resistor has a surface of 34~mm$^2$. The pulses are generated by a synthesizer at 200~MHz, driven by a function generator and are amplified by a RF amplifier able to provide more than 1~kW, with a gain around 64~dB (Bruker BLAH~1000). A directional coupler is used to control the  signal injected in the resistor and to measure the reflected power (which increases abruptly if the resistor burns). The whole setup is schematized in Figure~\ref{gene}.
The heater and sensor positions are shown in Figure~\ref{setup}. Two OSTs have been placed facing the heater. 

\subsection{Variation of the heat pulse power}

\begin{figure}[h]
\centering
\includegraphics[width=8cm]{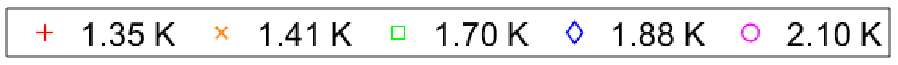}
\includegraphics[width=8.6cm]{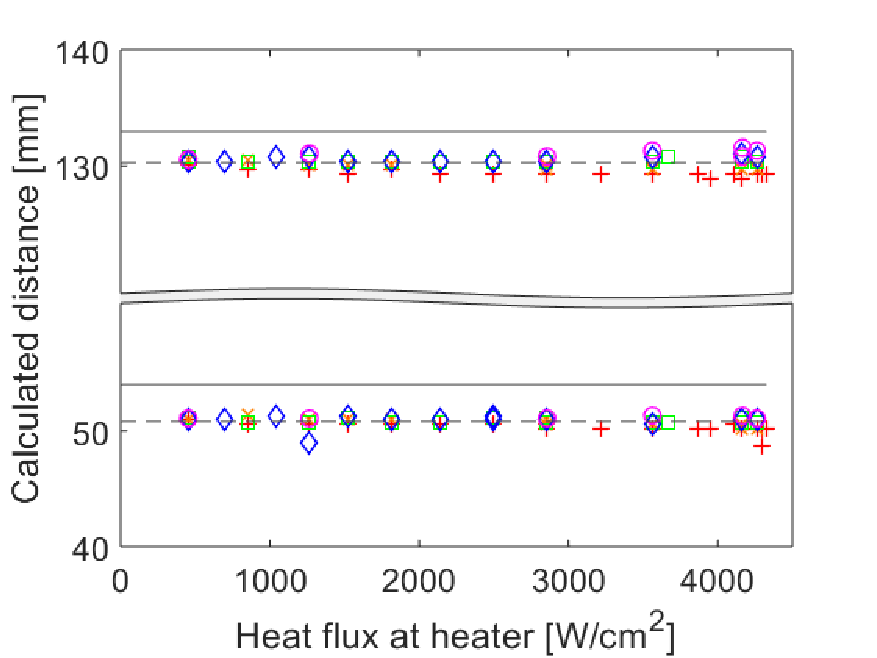}
\caption{Distances from heater to OSTs versus heat pulse power, calculated from times of arrival and recommended second sound velocity. For each OST, the average calculated values are shown (dashed lines) as well as the measured distances (solid lines).}
\label{powerT}
\end{figure}

For a set of temperatures in the range [1.3~-2.1~K] the influence of the heat pulse power level on the OSTs signals was observed. The pulse duration is kept equal to 100~\textmu s, and the power is increased from about 150~W to 1500~W, corresponding to a surface power from about 450 to~4500 W/cm$^2$. A second sound signal time of arrival has been measured for each OST, and the corresponding estimated distance from the heater to the OST has been calculated using the literature second sound velocities \cite{Donnelly}. These calculated distances have been plotted versus heat pulse power (Figure~\ref{powerT}), for both OSTs and for the whole set of temperatures. The results show clearly that the heat pulse power level does not have any influence on these calculated distances. For both OSTs, the mean calculated value has been plotted (dotted line), as well as the distance measured on the setup (solid line). As shown in Figure~\ref{setup}, these measured distances, resp. 54~mm and 133~mm, are from the heater to the closest point of each OST. In both cases, the mean calculated distances are around 3 mm lower than expected. This difference could be compatible with the inaccuracy of the distances measurements, due to the flexibility of the rods holding the OSTs. Indeed, it was observed that these rods were slightly deformed during the introduction of the setup in the cryostat.

\begin{figure}[h]
\centering
\includegraphics[width=8.6cm]{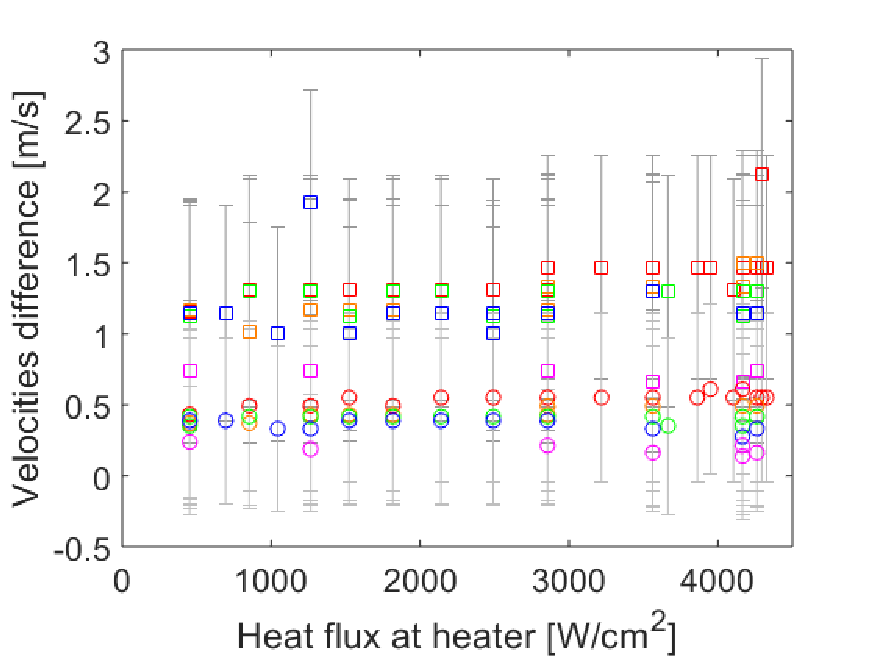}
\caption{Difference between the apparent second sound velocity and the literature values versus heat flux at heater. Values are given for OST~1 (squares) and OST~2 (circles), and for five sets of temperature: 1.35~K (red), 1.41~K (orange), 1.7~K (green), 1.88~K (blue) and 2.1~K (purple).}
\label{review1}
\end{figure}

However this distance difference can also be explained by a second sound velocity difference. This is shown in Figure \ref{review1} where the measured distances (54 mm and 133 mm) are divided by the times of arrivals of the OST signals ; the difference between this apparent second sound velocity and the literature values \cite{Donnelly} is plotted versus the heat flux at heater. One can observe that the apparent velocity is between 0.1 and 2 ms higher than expected, and that the difference is higher for OST1 than for OST2.
This results can be compared to the heat pulse experiments described in references \cite{Junginger2015} and \cite{Koettig2015}. The authors explore a lower level of heat flux than us: [0-700~W/cm$^2$] instead of [450-4500~W/cm$^2$]. Their results also present a higher than expected apparent velocity with a difference up to 1.5~m/s. In their measurements the velocity difference is dependent on the heating power below 200 W/cm$^2$. At higher heating powers the velocity difference remained constant. While this observation might explain slightly premature signals, this effect is too weak to explain the significant deviations that were observed in tests with cavities.

\subsection{Variation of the heat pulse energy}

\begin{figure}[h]
\centering
\includegraphics[width=8.6cm]{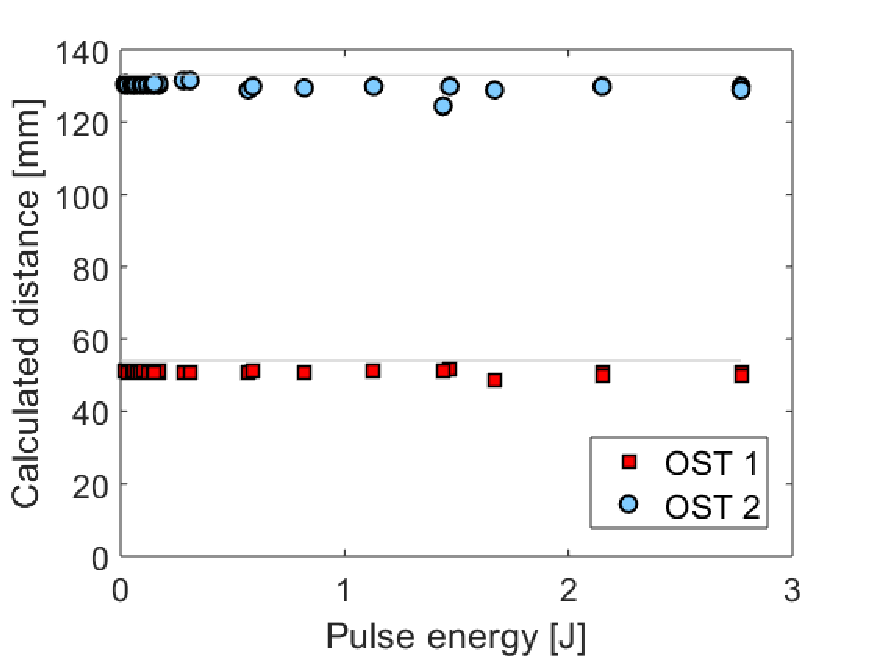}
\caption{Distances from heater to OSTs versus heat pulse energy, calculated from times of arrival and recommended second sound velocity. For each OST the measured distances is shown (solid lines).}
\label{powert}
\end{figure}

It is possible that the heat pulse energy, rather than the pulse power, could have an influence on the heat wave propagation. Experiments have been carried out at 1.7~K, where the pulse power was varied up to 1.5~kW, the pulse duration up to nearly 2~s and thus the pulse energy up to nearly 2.8~J. The results are presented in Figure~\ref{powert}, and show clearly that the pulse energy does not have any effect on the heat wave propagation time.

\subsection{Conclusions on heater experiments}

These experiments demonstrate that the heat pulse up to 1.5~kW and 2.8~J created heat waves fully compatible with the standard second sound waves in superfluid helium.

Thus the hypothesis of a heat wave propagating significantly faster in helium than the second sound waves has to be discarded, at least until the power and energy reached in these experiments. By comparison one can estimate the overall energy dissipated during a cavity quench, which corresponds to the energy stored in the cavity just before quench occurrence ($W_{stored}$). This quantity is related to the accelerating field reached in the cavity $E_{acc}$, the cavity frequency and some geometric parameters: the coupling coefficient $r/Q$ and the accelerating length $L_{acc}$\cite{Padam2008}.

\begin{equation}
W_{stored}=\frac{(E_{acc}L_{acc})^2}{\omega * r/Q} = 0.0141*E_{acc}^2
\end{equation}

All of the 1.3~GHz cavities tested have an $r/Q$ coefficient equal to 115~\textOmega~ and an accelerating length of 0.115~m. The maximum accelerating field varies from 30 to 37~MV/m (see Figure \ref{Q-Eacc}) and thus the maximal energy from 12.7 to 19.3~J. The time to dissipate this energy during the quench was not measured but assuming a typical quench duration in the order of one millisecond, one will obtain an average heating power of 12.7 to 19.3~kW. 

So the power and the energy dissipated in helium during a cavity quench is less than one order of magnitude higher than the maximal energy dissipated during the heater experiments. In contrast to the tests in the cavity no heat transfer through niobium did hinder the cooling. Taking this into account, the heating powers in the tests were even closer to the surface heat flux in the performed cavity tests. Even if the presented results cannot be directly transposed to cavity tests, at least they do not provide any argument to support the hypothesis of a heat wave propagating faster in helium than the standard second sound wave.

As additional information, during these experiments and for the highest levels of pulse energy, a cracking sound was audible, even if not recorded or measured. We identified this sound as the signature of a boiling as the highest surface heat flux were much higher than the reported literature data for the onset of boiling in superfluid helium \cite{VanSciver2012}. Actually, boiling phenomena in similar conditions have already been observed and recorded \cite{Peters2014}. Thus it seems that heat wave propagation is still compatible with the second sound velocity even if boiling in helium occurs.

\section{Propagation in Niobium}

As the origin of the premature signals could not be found to be caused by effects within the liquid helium, an influence of the cavity itself seems likely. Two effects come into mind to explain this. First, the quench propagation within in the cavity could be the reason. Second the heat transfer in the niobium itself could also cause this.

The quench of a cavity is generally caused by a local increase of the magnetic field on a small defect situated on the inner surface of the cavity. This spot is heated very rapidly, which causes the neighboring area to quench as well. This process continues until the heat input by the stored energy in the cavity decreases below the value of cooling by the surrounding niobium or helium. So the heat source on the outer surface of the cavity is rather an area, with a size depending on the dynamic heating and the dynamic heat diffusion in niobium and superfluid helium. Moreover, one can consider that the temperature increase due to the heat flux diffuses faster in the niobium than in helium, both due to the additional heating as well as the high thermal diffusivity of superconducting niobium. While both processes will certainly play a role, it is likely, that the rapid signal propagation will be dominated by the occurring quench phenomena.

\subsection{Probing fast signal propagation in niobium}

\begin{figure}[h]
\centering
\includegraphics[width=6cm]{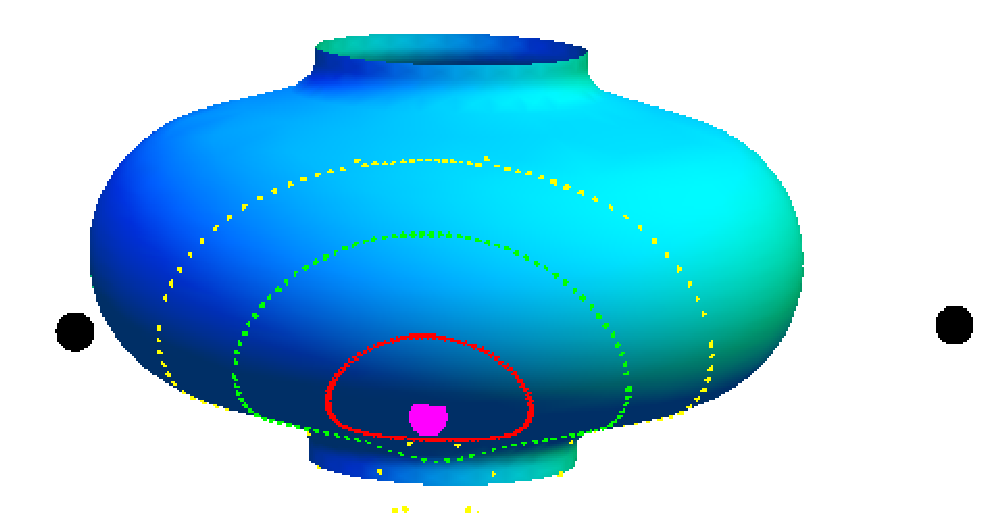}
\caption{Limit of the quench spot at different times (each time corresponds to a color). The large pink point is the quench position and the large dark points represent OSTs.}
\label{isotrope1}
\end{figure}

\begin{figure}[h]
\centering
\includegraphics[width=6cm]{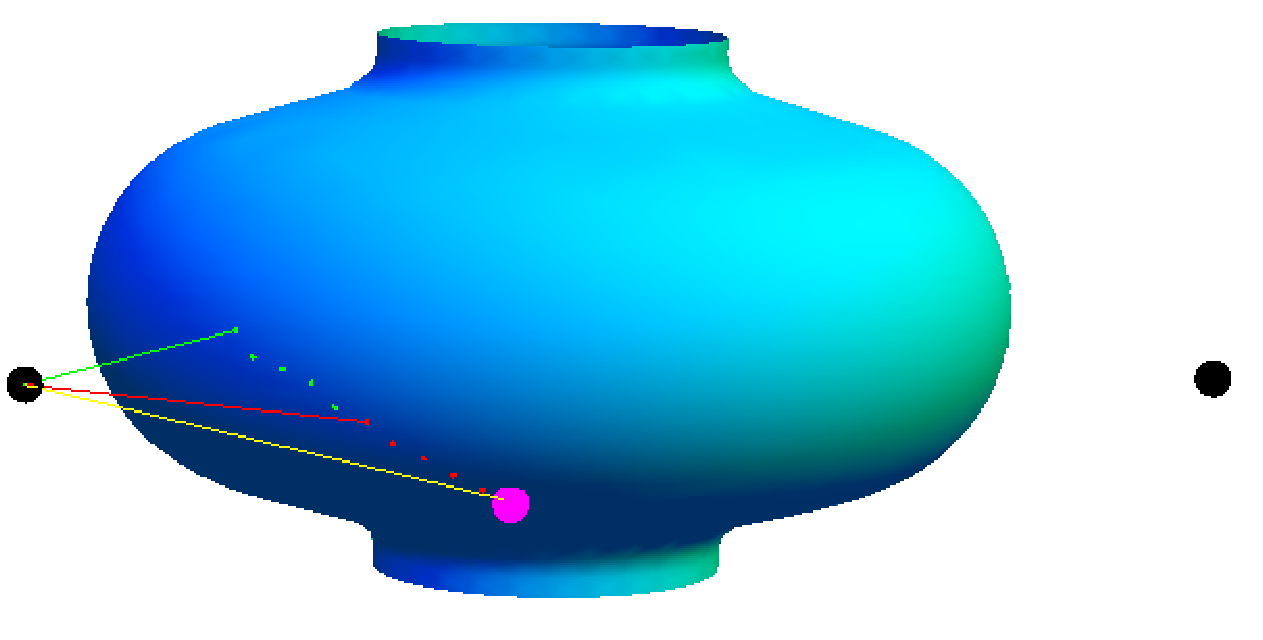}
\caption{Trajectory of the heat flux signal from quench spot (large pink point) to OST (large dark point) for different combination of heat transport in niobium and helium.}
\label{isotrope2}
\end{figure}

The signals arriving on an OST are the combination of second sound waves arriving directly from the quench position, and second sound emitted from other positions on the cavity. The very first signal arriving on the OST will be detected, but it could be ”emitted” by one of these positions rather than by the quench point itself. This situation has been discussed by Eichhorn et al. in a 2D case \cite{Eichhorn2015b}. It is illustrated in 3D in Figure \ref{isotrope1}, with an isotropic spread of the heat flux in the niobium. The heat spot limit is represented by the dotted points, each colored set corresponding to a given time after the beginning of the quench at the niobium surface. All the points of one color are situated at the same distance, on the cavity surface, from the quench point. Figure \ref{isotrope2} shows the different signals arriving on an OST, represented by a dark dot. The yellow line corresponds to a signal arriving directly from the quench location (large pink point) while the red and green lines show the signals arriving from two different points on the cavity surface.

To obtain the time of arrival, a minimization of the propagation in helium and in niobium has to be performed. This was implemented in Mathematica code. The velocity of the heat signal was fixed to 20~m/s in helium and to 70~m/s in niobium. Using a constant propagation velocity in niobium does not capture the underlying physical processes. These are highly dynamic and include many influencing factors, mainly the quench propagation and the thermal diffusion. Using a velocity to describe the processes in niobium was originally meant as an example calculation \cite{Peters2014}. It is based on simulated results in \cite{Liu2012}. This approach  was shown to match real systems surprisingly well as shown in \cite{Eichhorn2015b}.This is the reason, why the signal propagation in niobium is represented by this velocity.

\begin{figure}[h]
\centering
\includegraphics[width=8.6cm]{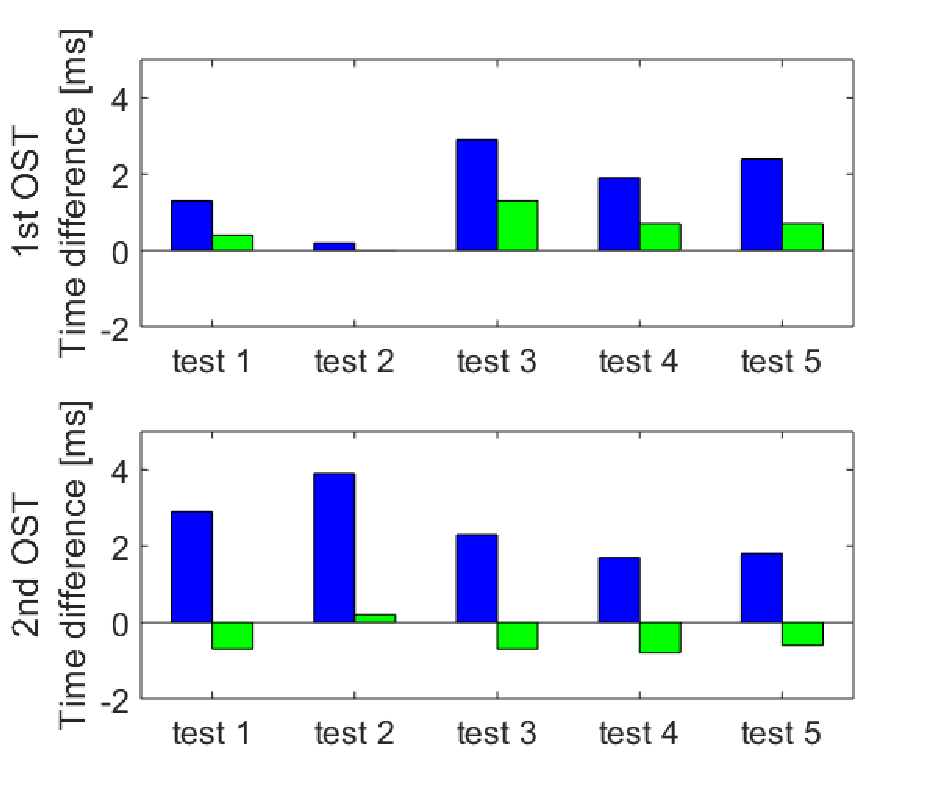}
\caption{Difference between calculated and measured  times of arrival of the 2nd sound signal on the two closest OSTs from the quench location at 1.7~K. Calculation with direct line of sight hypothesis (blue) and calculation with the mixed niobium/helium heat propagation model (green).}
\label{bar2}
\end{figure}

This minimization was applied to the geometry of the measured cavities. The results are given in Figure \ref{bar2}, together with a summary of the data from Figure \ref{bar1} giving results within direct line of sight propagation.

It can be seen that the calculation from this niobium-helium transport model reduces the error quite considerably for all measurements. It fits particularly well for cavity A with the measurements for the first OST and with an error less than 10\% for the second OST. In this case this model could describe the measured times of arrival. However, this model does not work as good for cavity~B, where the difference between measurements and calculation is still quite large. 

\begin{figure}[h]
\centering
\includegraphics[width=8.6cm]{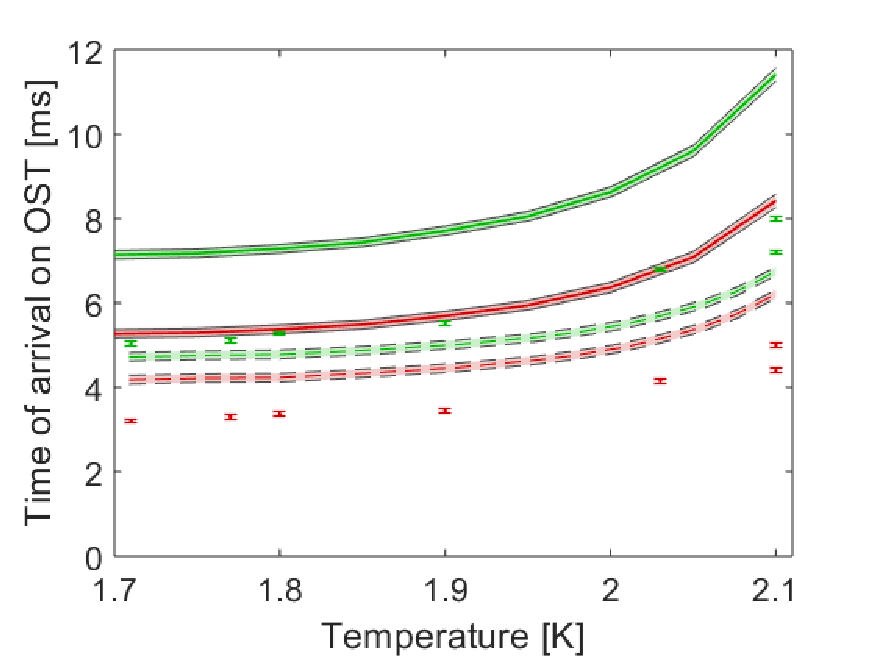}
\caption{Time of arrival on first (red) and second (green) OST with varying temperature for Test 4 (cavity B). Experimental data (points with error bars), calculated data with direct line of sight propagation (solid line) and with niobium-helium transport model (dashed line). The shaded area represent the error on calculation due to geometrical measurements uncertainties.}
\label{TvarOST2}
\end{figure}

\begin{figure}[h]
\centering
\includegraphics[width=8.6cm]{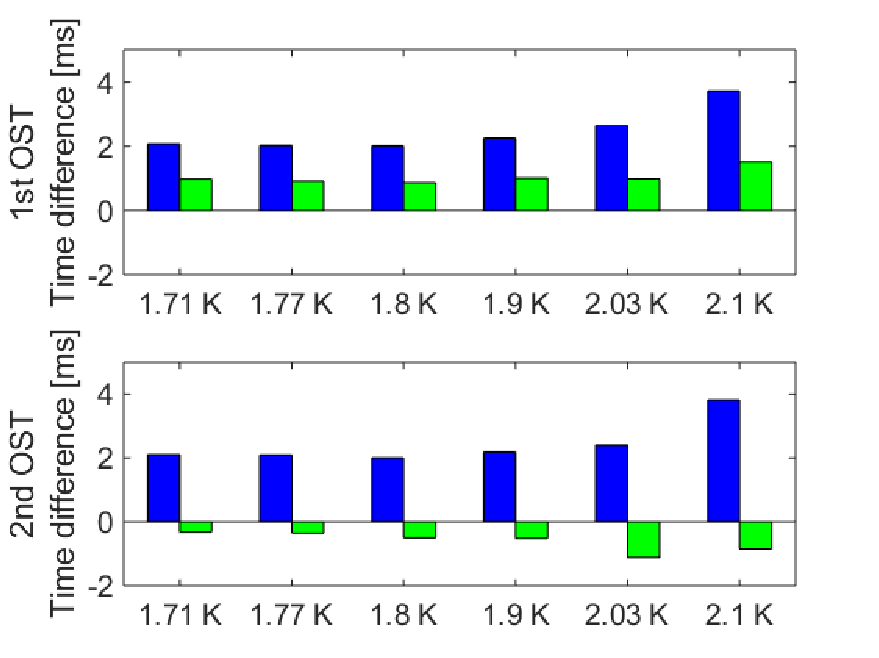}
\caption{Difference between calculated and measured  times of arrival of the 2\textsuperscript{nd} sound signal on the two closest OSTs from the quench location with varying temperature for Test 4. Calculation with direct line of sight propagation (blue) and calculation with the mixed niobium/helium heat propagation model (green). At 2.1~K the reference measured value is the average between the two real measured values shown in Figure} \ref{TvarOST2}
\label{bar3}
\end{figure}

Moreover this niobium-helium transport model has also been used for calculation with varying temperature and compared to the experimental results for test~4 with cavity~B. Results are shown in Figure \ref{TvarOST2} and Figure \ref{bar3}. In the whole temperature range of investigation this model reduces the error compared to the direct line of sight propagation calculations, but the differences keeps quite large: around 30\% for the 1\textsuperscript{st} OST and around -10\% for the 2\textsuperscript{nd} OST.

The difference between cavity~A and cavity~B could be explained by a number of reasons. The quench was located at different spots on the cavity. While in cavity A, the spot was on the equator, which is the region with the highest magnetic field on the inside, the spot on cavity B was in a lower field region (Figure \ref{carto}). The different spots will cause different patterns of quench propagation in the cavity. Another reason could be the different accelerating field in the cavities during the quench. In cavity B it is considerably higher, corresponding to an increase of 30\% of the dissipated energy for 34~MV/m and 50\% for 37~MV/m. These differences will lead to different average velocity of the temperature signal in niobium. As this velocity was just based on a number of assumptions, it is not surprising that the model is not an exact fit. Nevertheless this model gives an improved prediction of the time of arrival of the temperature signal. From these calculation it seems likely that the fast signal propagation in the niobium causes the premature signals.

\section{Conclusion}

In this paper the surprising effect of premature second sound signals in quench localization studies has been described and experimentally shown. This was confirmed by the combination of OST and temperature mapping measurements on two different cavities. These measurements allowed to test different ideas on the origin of this effect. The pure second sound propagation via the direct line of sight from a localized quench spot was used as a starting point. 

Two hypotheses were suggested to explain the early arrival of the second sound signal. A faster signal propagation in helium was probed in a dedicated experiment, where a heater with input power  up to $1.5$~kW, flux power to $4.3$~kW/cm$^2$ and energy $2.8$~J was used to simulate the quench spot. Such amounts of power of energy had never been reached in similar measurements. However, no faster signal propagation could be found under these conditions, which are lower than in actual cavities, but in a relevant regime. Propagation in niobium was probed by a model, which was used to evaluate the quench tests with the cavities. This model was based on a number of assumptions. Although these assumptions were quite broad, a considerable improvement on the prediction of the time of arrival for the second sound at the OSTs could be obtained. It can be concluded, that while a fast signal propagation in helium seems very unlikely, the processes in the niobium are likely to induce the described effect. It is assumed that the quench propagation on the inner cavity wall and the thermal diffusion in the niobium are the main cause, in the case of quenching cavities.

%\section{References}

\end{document}